\documentclass[12pt]{article}

\usepackage{geometry}
\usepackage{amsmath}
\usepackage{amssymb}

\newcommand{\al}{\alpha}

\newcommand{\de}{\delta}

\newcommand{\thet}{\theta}

\newcommand{\la}{\lambda}

\newcommand{\pa}{\partial}

\newcommand{\lp}{\left(}
\newcommand{\rp}{\right)}
\newcommand{\li}{\left<}
\newcommand{\ri}{\right>}

\newcommand{\fr}[2]{\frac{#1}{#2}}

\begin{document}

\title{Quantum mechanics from a universal action reservoir}
\author{A. Garrett Lisi\thanks{alisi@hawaii.edu}}
\maketitle

\begin{abstract}
A heuristic derivation of quantum mechanics using information theory requires a foundational physical principle: the existence of a universal action reservoir, analogous to the energy reservoir of a canonical ensemble.
\end{abstract}

\section{Introduction}

Modern information theory provides a practical framework for mathematically inferring probabilities for possible occurrences based on what is known by an observer.\cite{Shan} This Bayesian approach -- associating probabilities with likelihoods for a system rather than with frequencies of events -- works for simple systems and single particles as well as for aggregates of many. The methods provide a new perspective on constructions in statistical mechanics and, along with one necessary physical principle, lead directly to quantum mechanics.

\section{Quantum Mechanics}

An important clue towards establishing a good conceptual foundation for quantum mechanics is that so much can be done without it. Even though the universe is fundamentally quantum, classical descriptions match most of our macroscopic experience. Also, on the technical level, most descriptions of classical systems can be successfully quantized to obtain quantum descriptions. This seems strange from the perspective of taking QM to be fundamental -- it is as if quantum mechanics is something extra, rather than something different. Finally, the success of special and general relativity imply QM should be formulated relativisticly, without a special role for time. These facts, and its use in QFT, indicate Feynman's path integral formulation\cite{Feyn} is the best foundational description, summarized by the use of the quantum partition function,
$$ Z = \sum_{paths} e^{-\fr{1}{i \hbar} S[path]}$$
The sum (a path integral) is over all paths (or histories) in classical configuration space, and $S$ is the classical action associated with each path. The formal similarity to the partition function for a statistical canonical ensemble is striking, 
$$ Z = \sum_{states} e^{-\fr{1}{k_B T} E[state]}$$
and the two follow from identical derivations. Since path integrals (as casually treated here) may not always be mathematically sensible, the following derivation of quantum mechanics is best viewed as a heuristic argument -- primarily serving to uncover the physical principles and interpretations underlying the theory. 

Consider a system described by a set of configuration variables, $q$. A $path$ of the the system (analogous to a canonical $state$) is a continuous function, $path = q(t)$, parameterized by a set of one or more parameters, $t$. For a physical system, an action, $S[path] = S[q(t)] \in \mathbb{C}$ (possibly complex), is associated with each $path$. Typically, the action is an integral of a system Lagrangian over the parameters, $S = \int dt \, L(q,\dot{q})$. Every observer naturally associates a probability, $p[path] = p[q(t)]$ (abbreviated $p[q]$), with each possible path of the system. The self-information is the negative logarithm of the probability for the path, $\log{\fr{1}{p[path]}} = - \log{p[path]}$ -- it is the amount of information the observer would get by knowing the system is on $path$. The entropy (synonymous with ignorance, Shannon uncertainty, and information entropy) of a probability distribution is
\begin{equation} H = - \sum_{paths} p[path] \log{p[path]} = - \int{Dq \, p[q] \log{p[q]}} \label{ent} \end{equation}
in which the path integral appears as a sum over system paths. This number represents the observer's overall ignorance of which path the system takes. The principle of maximum entropy is the reasonable assertion that the ignorance of the probability distribution should be maximized, constrained by what the observer knows. One such constraint is that all the probabilities should sum to one,
$$1 = \sum_{paths} p[path] = \int{Dq \, p[q]}$$
consistent with the fact the system is known to exist. For the case of a statistical (thermodynamic) canonical ensemble, a system is presumed to be in equilibrium with an energy reservoir of known temperature, implying an average state energy. Analogously, quantum mechanics appears to derive from a single physical principle:
\begin{quote}
{\bf A quantum system is one in contact with a universal action reservoir, providing a known expected path action.}
\end{quote}
$$\overline{S} = \li S \ri= \sum_{paths} p[path] \, S[path] = \int{Dq \, p[q] S[q] } \in \mathbb{C}$$
Maximizing the entropy subject to these two constraints gives the correct probability distribution. Employing Lagrange multipliers, $\la \in \mathbb{C}$ and $\al \in \mathbb{C}$, the effective entropy to be extremized is
\begin{align*}
H' & = - \int{Dq \, p[q] \log{p[q]}} + \la \lp 1 - \int{Dq \, p[q]} \rp + \al \lp \overline{S} - \int{Dq \, p[q] S[q] } \rp \\
 & = \la + \al \overline{S} - \int{Dq \, \lp p[q] \log{p[q]} + \la p[q] + \al p[q] S[q] \rp}
\end{align*} 
Varying the probability distribution gives
$$\de H' = - \int{Dq \, \lp \de p[q] \rp \lp \log{p[q]} + 1 + \la + \al S[q] \rp}$$
which is extremized when $\de H' = 0$, corresponding to the probability distribution,
\begin{equation} p[q] = e^{-1-\la}e^{-\al S[q]} = \fr{1}{Z} e^{-\al S[q]} \label{prob} \end{equation}
compatible with the knowledge constraints. Varying the Lagrange multipliers enforces the two constraints, giving $\la$ and $\al$. Specifically, $e^{-1-\la} = \fr{1}{Z}$, in which the quantum partition function is
$$ Z = \int{Dq \, e^{-\al S[q]}}$$
while $\al$ is determined by solving
$$ \overline{S} = \int{Dq \, S[q] p[q] } = \fr{1}{Z} \int{Dq \, S[q] e^{-\al S[q]} } = - \fr{\pa}{\pa \al} \log Z $$
The resulting Lagrange multiplier value, $\al = \fr{1}{i \hbar}$, is an intrinsic quantum variable directly related to the average path action, $\overline{S}$, of the universal reservoir. Planck's constant is analogous to the thermodynamic temperature of a canonical ensemble, $i \hbar \leftrightarrow k_B T$. Being constant reflects its universal nature -- analogous to an isothermal canonical ensemble. If allowing $\al$, $\overline{S}$, and many other numbers to be imaginary is unpalatable, an alternative is to Wick rotate to real quantities (analytic continuation) in a parameter, $t \rightarrow e^{i \thet} t$ with $0 \leq \thet \leq \fr{\pi}{2}$. Everything knowable of the system is determined using the probability distribution and how it changes when system parameters are varied. All the tools of statistical mechanics and thermodynamics are available. The expected value for any functional of system path is
$$\li F \ri = \sum_{paths} F(q[path]) \, p[path] = \int{Dq \, F(q) \, p[q]} = \fr{1}{Z} \int{Dq \, F(q) \, e^{-\al S[q]}}$$
The probability for the system path to be among a set of possibilities is found by summing the probabilities of paths in the set,
$$p(set) = \sum_{paths} \de^{set}_{path} \, p[path] = \int{Dq \, \de(set - q) \, p[q]}$$
Typically, the system action reverses sign under inversion of the parameter integration limits,
$$S^{t'} = \int^{t'} dt \, L(q,\dot{q}) = - \int_{t'} dt \, L(q,\dot{q}) = - S_{t'} $$
This implies the probability for the system path to pass through configuration $q'$ at parameter value $t'$ is
$$p(q',t') = \int{Dq \, \de(q'(t') - q) \, p[q]} = \lp \int^{q(t')=q'}{Dq \, p^{t'}[q]} \rp \lp \int_{q(t') = q'}{Dq \, p_{t'}[q]} \rp = \Psi(q',t') \, \Psi^*(q',t')$$
in which
$$\Psi(q',t') = \int^{q(t')=q'}{Dq \, p^{t'}[q]} = \fr{1}{\sqrt{Z}} \int^{q(t')=q'}{Dq \, e^{ - \al S^{t'}}}$$
The quantum wavefunction, $\Psi(q',t')$, is the complex amplitude of paths with $t < t'$ meeting at $q'$, while its complex conjugate, $\Psi^*(q',t')$, is the amplitude of paths with $t > t'$ leaving from $q'$ -- these probability amplitudes multiply to give the probability of the system passing through $q'(t')$. (This wavefunction description is subordinate to the probability distribution \eqref{prob}, and works only when $t'$ is a physical parameter and the system is $t'$ symmetric, providing a real partition function, $Z$.) If new information is discovered about the system, the probability distribution must be adjusted accordingly by maximizing the entropy \eqref{ent} within the bounds of the new knowledge constraints -- ``wavefunction collapse.'' Also, just as in thermodynamic systems, the quantum system is never known with certainty to be on a single $path$. (The insistence that the system does follow one path, and we just don't know what it is, is the Bohmian interpretation.)

\section{Discussion}

The practical use of path integrals, partition functions, and Wick rotation is well established in quantum field theory. Nevertheless, the woolly nature of the path integral (as treated here) implies the presented derivation qualifies only as a heuristic sketch. Also, the physical interpretation and justification for Wick rotation and complex probability distributions remains an open question. It is not yet clear (to the author) whether this path likelihood approach to quantum mechanics can be made computationally sound and lead to new physical predictions, but it seems worth investigating. The main product of the work is the proposal of a new physical principle for the foundation and interpretation of quantum mechanics: a universal background action. Additionally, an observer dependent probability distribution is compatible with the Relational Quantum Mechanics interpretation and the emergence of thermal time from a probability distribution.\cite{Rove,Conn} This point of view gives wavefunction collapse the trivial interpretation of likelihood adjustment based on knowledge acquisition. Finally, the relativistic nature of this background action formulation gives some hope it may find application in a viable theory of quantum gravity.

\end{document}